\documentclass[aps,pre,twocolumn,showkeys,superscriptaddress]{revtex4}
\usepackage{epsfig}
\usepackage{times}
\begin{document}

\title{Antisocial pool rewarding does not deter public cooperation}

\author{Attila Szolnoki}
\email{szolnoki@mfa.kfki.hu}
\affiliation{Institute of Technical Physics and Materials Science, Centre for Energy Research, Hungarian Academy of Sciences, P.O. Box 49, H-1525 Budapest, Hungary}

\author{Matja{\v z} Perc}
\email{matjaz.perc@uni-mb.si}
\affiliation{Faculty of Natural Sciences and Mathematics, University of Maribor, Koro{\v s}ka cesta 160, SI-2000 Maribor, Slovenia}
\affiliation{Department of Physics, Faculty of Sciences, King Abdulaziz University, Jeddah, Saudi Arabia}
\affiliation{CAMTP -- Center for Applied Mathematics and Theoretical Physics, University of Maribor, Krekova 2, SI-2000 Maribor, Slovenia}

\begin{abstract}
Rewarding cooperation is in many ways expected behaviour from social players. However, strategies that promote antisocial behaviour are also surprisingly common, not just in human societies, but also among eusocial insects and bacteria. Examples include sanctioning of individuals who behave prosocially, or rewarding of freeriders who do not contribute to collective enterprises. We therefore study the public goods game with antisocial and prosocial pool rewarding in order to determine the potential negative consequences on the effectiveness of positive incentives to promote cooperation. Contrary to a naive expectation, we show that the ability of defectors to distribute rewards to their like does not deter public cooperation as long as cooperators are able to do the same. Even in the presence of antisocial rewarding the spatial selection for cooperation in evolutionary social dilemmas is enhanced. Since the administration of rewards to either strategy requires a considerable degree of aggregation, cooperators can enjoy the benefits of their prosocial contributions as well as the corresponding rewards. Defectors when aggregated, on the other hand, can enjoy antisocial rewards, but due to their lack of contributions to the public good they ultimately succumb to their inherent inability to secure a sustainable future. Strategies that facilitate the aggregation of akin players, even if they seek to promote antisocial behaviour, thus always enhance the long-term benefits of cooperation.
\end{abstract}

\maketitle

\section*{1. Introduction}
The sustainability of modern human societies relies on cooperation among unrelated individuals \cite{ostrom_90}. Situations that require cooperative behaviour for socially beneficial outcomes abound and range from taxpaying and voting to neighbourhood watch, recycling, and climate change mitigation \cite{wang_rw_jrsif11, rand_tcs13, wu_zx_epl15, pacheco_plrev14, rong_njp15}. The crux of the problem lies in the fact that, while cooperation leads to group-beneficial outcomes, it is jeopardized by selfish incentives to free-ride on the contributions of others. Excessive short-term benefits to individuals who act as selfish maximizers create systemic risks that may nullify the long-term benefits of cooperation and lead to  the tragedy of the commons \cite{hardin_g_s68}. Fortunately, we have strong predispositions to behave morally even when this in conflict with our material interests \cite{nowak_11}. The innate human drive to act prosocially is a product of our evolution as a species, as well as our unique capacity to internalise norms of social behaviour \cite{hrdy_11, bowles_11}. Yet, it is also important to note that impaired recognition and absent cognitive skills are likewise potential triggers of antisocial rewarding, in particular since under such circumstances the donor of the reward is likely to be unable to distinguish between cheaters and cooperators. As such, the concepts of mutualism and second-order free-riding are by no means limited to human societies, but apply just as well to certain eusocial insects as well as to bacterial societies \cite{he_jz_pone14}.

Despite favourable predispositions, however, cooperation is often subject to both positive and negative incentives \cite{andreoni_aer03, szolnoki_prx13, okada_ploscb15, rand_pone15, he_jz_srep15}. Positive incentives typically entail rewards for behaving prosocially \cite{dreber_n08, hilbe_prsb10, hauert_jtb10, szolnoki_epl10, szolnoki_njp12}, while negative incentives typically entail punishing free-riding \cite{fehr_aer00, gardner_a_an04, henrich_s06b, sigmund_tee07, raihani_tee12, rand_ncom11, powers_jtb12, hauser_jtb14, mccabe_14}. However, just like public cooperation incurs a cost for the wellbeing of the common good, so does the provisioning of rewards or sanctions incur a cost for the benefit or harm of the recipients. Individuals that abstain from dispensing such incentives therefore become second-order freeriders \cite{fehr_n04}, and they are widely believed to be amongst the biggest impediments to the evolutionary stability of rewarding and punishing \cite{panchanathan_n04, hauert_s07, helbing_ploscb10, hilbe_srep12, chen_xj_njp14}.

In addition to being costly, the success of positive and negative incentives is challenged by the fact that they can be applied to promote antisocial behaviour. Antisocial punishment, that is, the sanctioning of group members who behave prosocially, is widespread across human societies \cite{herrmann_s08}. Moreover, antisocial rewarding is present in various inter-specific social systems, where the host often rewards the parasitic species of a symbiont \cite{wang_rw_e10}. This phenomenon is due to the inability of the donor to distinguish defectors and cooperators. Recent theoretical work also indicates that antisocial punishment can prevent the coevolution of punishment and cooperation \cite{rand_jtb10}, just like antisocial rewarding can lead to the breakdown of cooperation if the latter is contingent on pool rewarding \cite{dos-santos_m_prsb15}. In theory, the resolution of such social traps involves rather complex set-ups, entailing the ability of second-order sanctioning, elevated levels of effectiveness of prosocial incentives in comparison to antisocial incentives, or the decreased ability to dispense antisocial incentives due to the limited production of public goods in environments with low levels of cooperation.

Here we study what happens if both competing strategies are able to invest into a rewarding pool to support akin players. How does such a strategy-neutral intervention influence the evolutionary outcome of a public goods game? We consider a four-strategy game, where beside traditional cooperators and defectors also rewarding cooperators and rewarding defectors are present. Rewarding cooperators reward other rewarding cooperators, while rewarding defectors reward other rewarding defectors, thus representing prosocial and antisocial pool rewarding, respectively. Noteworthy, our setup differs slightly from a recently studied model where rewarding players could be utilized directly by non-rewarding competitors \cite{dos-santos_m_prsb15}. In our case, however, we focus on the impact of the strategy-neutral intervention in the form of pool rewarding. In addition to the well-mixed game, we mainly study the game in a structured population, where everybody does not interact with everybody else, and the interactions that do exist are not random \cite{wasserman_94, christakis_09, apicella_n12}. The importance of structured populations for the outcome of evolutionary social dilemmas was reported first by Nowak and May \cite{nowak_n92b}, and today the positive effects of spatial structure on the evolution of cooperation are well-known as network reciprocity \cite{nowak_s06, rand_pnas14}. Several recent reviews are devoted to evolutionary games in structured populations \cite{szabo_pr07, perc_bs10, roca_plr09, santos_jtb12, perc_jrsi13, szolnoki_jrsif14, wang_z_epjb15}.

The consideration of prosocial and antisocial pool rewarding in structured populations is thus an important step that promises to elevate our understanding of the impact of strategies that aim to promote antisocial behaviour in evolutionary games. As we will show, antisocial rewarding does not hinder the evolution of cooperation from a random state in structured populations, and in conjunction with prosocial rewarding, it still has positive consequences in that it promotes the spatial selection for cooperation in evolutionary social dilemmas. This counterintuitive outcome can be understood through pattern formation that facilitates the aggregation of players who adopt the same strategies, which in turn helps to reveal the long-term benefits of cooperation in structured populations.

\section*{2. Material and methods}
The public goods game is a stylized model of situations that require cooperation to achieve socially beneficial outcomes despite obvious incentives to free-ride on the efforts of others. We suppose that players form groups of size $G=5$, where they either contribute $c = 1$ or nothing to the common pool. After the sum of all contributions is multiplied by the synergy factor $r_1>1$, the resulting public goods are distributed equally amongst all the group members irrespective of their contribution to the common pool. In parallel to this traditional version of the public goods game entailing cooperators ($C$) and defectors ($D$), two additional strategies run an independent pool rewarding scheme. These are rewarding cooperators ($R_C$) and rewarding defectors ($R_D$), who essentially establish a union-like support to aid akin players. Accordingly, rewarding cooperators contribute $c=1$ to the prosocial rewarding pool. The sum of all contributions in this pool is subsequently multiplied by the synergy factor $r_2>1$, and the resulting amount is distributed equally amongst all $R_C$ players in the group. Likewise, at each instance of the public goods game all rewarding defectors contribute $c=1$ to the antisocial rewarding pool. The sum of all contributions in this pool is subsequently multiplied by the same synergy factor $r_2>1$ that applies to the prosocial rewarding pool, and the resulting amount is distributed equally amongst all $R_D$ players in the group. We are thus focusing on the consequences of union-like support to akin players, without considering second-order free-riding. It is therefore important that
we consider strategy-neutral pool rewarding in that individual contributions to the prosocial and the antisocial rewarding pool are the same ($c=1$), as is the multiplication factor $r_2$ that is subsequently applied. Otherwise, if an obvious disadvantage would be given to either the prosocial or the antisocial rewarding pool, the outcome of the game would become predictable. We also emphasize that, in order to consider the synergistic consequence of mutual efforts and to avoid self-rewarding of a lonely player \cite{brandt_pnas06}, we always apply $r_2=1$ if only a single individual contributed to the rewarding pool.

In addition to the well-mixed version of the game, we primarily consider the spatial game. We emphasize that the importance of a structured population is not restricted to human societies, but applies just as well to bacterial societies, where the interaction range is typically limited, especially in biofilms and in vitro experiments \cite{drescher_cb14, kerr_n02}. Biological mechanisms that are responsible for the population being structured rather then well-mixed typically include limited mobility, time and energy constrains, as well as cognitive preferences in humans and higher mammals. In the corresponding model, the public goods game is staged on a square lattice with periodic boundary conditions where $L^2$ players are arranged into overlapping groups of size $G=5$, such that everyone is connected to its $G-1$ nearest neighbours. Accordingly, each individual belongs to $g=1,\ldots,G$ different groups. The square lattice is the simplest of networks that allows us to take into account the fact that the interactions among humans are inherently structured rather than well-mixed or random. Despite of its simplicity, however, there exist ample evidence in support of the fact that the square lattice suffices to reveal all the feasible evolutionary outcomes for games that are governed by group interactions \cite{szolnoki_pre09c, szolnoki_pre11c}, and also that these outcomes are qualitatively independent of the details of the interaction structure \cite{perc_jrsi13}. As an alternative, and to explore the robustness of our findings, we nevertheless also consider regular small-world networks, where a fraction $Q$ of all links is randomly rewired once before the start of the game \cite{szabo_jpa04}.

The considered evolutionary game in a structured population is studied by means of
Monte Carlo simulations, which are carried out as follows. Initially each player on site $x$ is designated either as a cooperator, defector, rewarding cooperator, or a rewarding defector with equal probability. Next, the following elementary steps are iterated repeatedly until a stationary solution is obtained. A randomly selected player $x$ plays the public goods game with its $G-1$ partners as a member of all the $g=1,\ldots,G$ groups, whereby its overall payoff $\Pi_{s_x}$ is thus the sum of all the payoffs $\Pi_{s_x}^{g}$ acquired in each individual group as described in the preceding subsection. Next, player $x$ chooses one of its nearest neighbours at random, and the chosen co-player $y$ also acquires its payoff $\Pi_{s_y}$ in the same way. Finally, player $x$ enforces its strategy $s_x$ onto player $y$ with a probability given by the Fermi function $w(s_x \to s_y)=1/\{1+\exp[(\Pi_{s_y}-\Pi_{s_x}) /K]\}$, where $K=0.5$ quantifies the uncertainty by strategy adoptions \cite{szolnoki_pre09c}, implying that better performing players are readily adopted, although it is not impossible to adopt the strategy of a player performing worse. Such errors in decision making can be attributed to mistakes and external influences that adversely affect the evaluation of the opponent. Each full Monte Carlo step (MCS) gives a chance to every player to enforce its strategy onto one of the neighbours once on average.

The average fractions of cooperators ($f_{C}$), defectors ($f_{D}$), rewarding cooperators ($f_{R_C}$), and rewarding defectors ($f_{R_D}$) on the square lattice were determined in the stationary state after a sufficiently long relaxation time. Depending on the proximity to phase transition points and the typical size of emerging spatial patterns, the linear system size was varied from $L=400$ to $1200$, and the relaxation time was varied from $10^4$ to $10^5$ MCS to ensure that the statistical error is comparable with the line thickness in the figures.

\section*{3. Results}

\subsection*{(a) Evolution in a well-mixed population}
From the pairwise comparison of strategies it follows that pool rewarding is dominant. Accordingly, the original 4-strategy game can be reduced to a 2-strategy game, where the $R_C$ and $R_D$ strategies compete. Designating by
$n_{R_C}$ the number of rewarding cooperators and by $n_{R_D}$ the number of rewarding defectors among other players in a group, the payoffs of the two competing strategies are
\begin{eqnarray}
\Pi_{R_D} &=& r_1 \frac{n_{R_C}}{G}+r_2 \delta(R_D) - 1\,,\\
\Pi_{R_C} &=& r_1 \frac{n_{R_C}+1}{G} - 1 + r_2 \delta(R_C) - 1\,,
\end{eqnarray}
where
\begin{equation}
\delta(s) = \left\{
\begin{array}{cl}
1& \textrm{if } n_s > 0\\
\frac{1}{r_2}& \textrm{otherwise}\,.
\end{array}
\right.
\end{equation}
By designating the fraction of $R_C$ players as $x$, the corresponding replicator equation becomes
\begin{equation}
\dot{x} = x [P_{R_C} - (x P_{R_C}+(1-x) P_{R_D}) ]\,.
\label{replica}
\end{equation}
Here
\begin{equation}
P_s = \sum_{n_{R_C},n_{R_D}} \frac{(G-1)!}{n_{R_C}!n_{R_D}!} x^{n_{R_C}} (1-x)^{n_{R_D}} \Pi_s\,,
\end{equation}
where $0\le n_s \le G-1$ and $\sum n_s = G-1$ are always fulfilled.

Starting from a random initial state, where both competing strategies are equally common ($x_i=0.5$), the solution of Eq.~\ref{replica} indicates that the population will always terminate into the full $R_D$ state if $r_1 < G$, and this independently of the value of $r_2$. In other words, the introduction of strategy neutral rewards cannot help cooperators if they are not already predominant in the initial population. Accordingly, the introduced rewards will not avert from the tragedy of the commons when the competing strategies start the evolutionary game equally strong.

\begin{figure}
\centerline{\epsfig{file=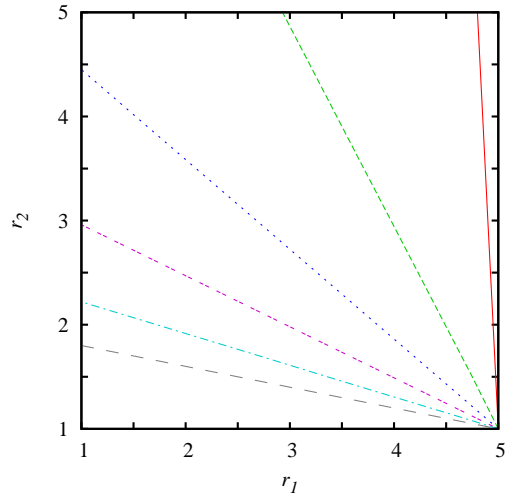,width=7cm}}
\caption{In a well-mixed population, when both competing strategies are initially equally common, the extinction of rewarding cooperators is unavoidable if $r_1 < G$, and this independently of the value of $r_2$. However, if $R_C$ players are initially in the majority, then a new stable state emerges, where rewarding cooperators are the only players remaining in the population. In this bistable case, the border of the attractive basin depends sensitively on the initial fraction $x_i$ of $R_C$ players. Lines in the figure show the border of the two basins, as obtained for $x_i=0.51$, $0.6$, $0.7$, $0.8$ and $0.9$, from top to bottom on the $r_1 - r_2$ parameter plane. The bottom-most dashed gray line shows the border in the limiting case, when there is an infinitesimally small minority of rewarding defectors initially present in the well-mixed population.}
\label{border}
\end{figure}

However, if $R_C$ players are somehow able to aggregate, then a significantly new situation emerges. This condition can be reached by assuming $x_i>0.5$, when rewarding cooperators form the majority in the initial population. In this case, the full $R_C$ and the full $R_D$ state becomes an attractor point, but the border of their basins depends sensitively on the values of $x_i$. This effect is illustrated in Fig.~\ref{border}, where we have plotted the border of the two stable solutions on the $r_1 - r_2$ parameter plane.

\subsection*{(b) Evolution in a structured population}

\begin{figure}
\centerline{\epsfig{file=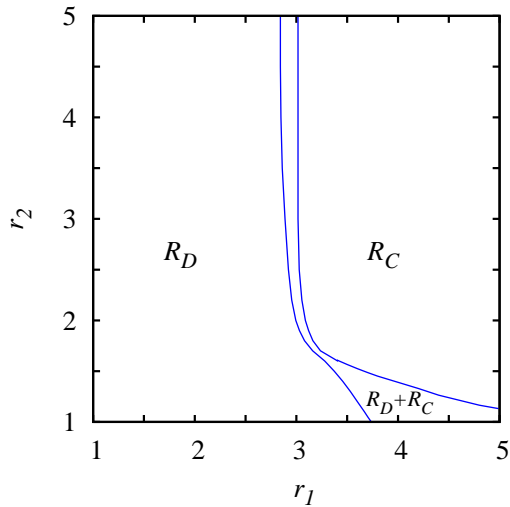,width=7cm}}
\caption{Phase diagram of the studied spatial public goods game, demonstrating that the presence of antisocial rewarding does not hinder prosocial rewarding to promote cooperation. Depicted are strategies that remain on the square lattice after sufficiently long relaxation times as a function of the multiplication factor for the public goods pool $r_1$ and the multiplication factor for the antisocial and prosocial rewarding pool $r_2$. Solid blue lines denote continuous phase transitions. Neither cooperators ($C$) nor defectors ($D$) who abstain from participating in pool rewarding are able to survive in the stationary state. Instead, for low values of $r_1$ rewarding defectors ($R_D$) dominate, while for sufficiently high values of $r_1$ and $r_2$ rewarding cooperators ($R_C$) prevail. In-between is a rather narrow two-strategy $R_D+R_C$ phase, where both rewarding strategies coexist. Interestingly, for example at $r_1=3.5$, increasing solely the value of $r_2$ can lead the population from a pure $R_D$ to a pure $R_C$ phase, thus indicating clearly that rewarding, even if applied to both strategies, still promotes cooperation.}
\label{phase}
\end{figure}

The lesson learned from the preceding subsection is that rewarding cooperators should initially constitute the majority of the population to survive. Otherwise, if their strength in numbers is absent, rewarding defectors inevitably take over. In a structured population, however, this special initial condition can spontaneously emerge locally, during the course of evolution, without there being an obvious advantage given to rewarding cooperators at the outset. The fundamental question then is whether such a positive local solution is viable and able to spread across the whole population, or rather if it is unstable and folds back to the defector-dominated state. To clarify this, we perform systematic Monte Carlo simulations to obtain the phase diagram for the whole $r_1-r_2$ parameter plane, as shown in Fig.~\ref{phase}. Before addressing the details, we emphasize that the reported stationary states are highly stable and fully independent of the initial conditions, which is a fundamental difference from the well-mixed solutions we have reported above. Starting with the $r_2=1$ line, which implies the absence of pool rewarding, we note that cooperators survive only if the critical value of $r_1$ is $r_{1_c}>3.74$ \cite{szolnoki_pre09c}. The fact that this value is still lower than the group size $G=5$, which would be the threshold in a well-mixed population, is due to network reciprocity. The latter enables cooperators to form compact clusters and so protect themselves against being wiped out by defectors \cite{nowak_n92b}. Taking this as a reference value, we can appreciate at a glance that, even in the presence of antisocial rewarding, prosocial rewarding still promotes the evolution of cooperation. However, neither defectors ($D$) nor cooperators ($C$) who abstain from pool rewarding can survive if $r_2>1$. Indeed, as in the well-mixed case, only rewarding defectors ($R_D$) and rewarding cooperators ($R_C$) remain in the stationary state, depending on the value of $r_1$ and $r_2$. This outcome can be understood since players that do engage in pool rewarding collect payoffs that exceed their initial contributions to the rewarding pool.

In terms of the relation between $R_D$ and $R_C$ players, it is interesting to note that the introduction of strategy-neutral pool rewarding unambiguously supports the cooperative strategy. In particular, as we increase the value of $r_2$ and thus increase also the efficiency of rewarding, the critical value of $r_1$ where $R_C$ players are able to survive decreases steadily. Likewise decreasing is the $r_1$ threshold for complete dominance of the $R_D$ strategy. At specific values of $r_1$, for example at $r_1=3.5$, it is even possible to go from the pure $R_D$ phase to the pure $R_C$ phase solely by increasing the value of $r_2$. Thus indeed, even if the prosocial pool rewarding scheme is accompanied by an equally effective antisocial pool rewarding scheme, in structured populations the evolution of cooperation from a neutral or even from an adverse initial state is still promoted well past the boundaries imposed by network reciprocity alone.

\begin{figure*}
\centerline{\epsfig{file=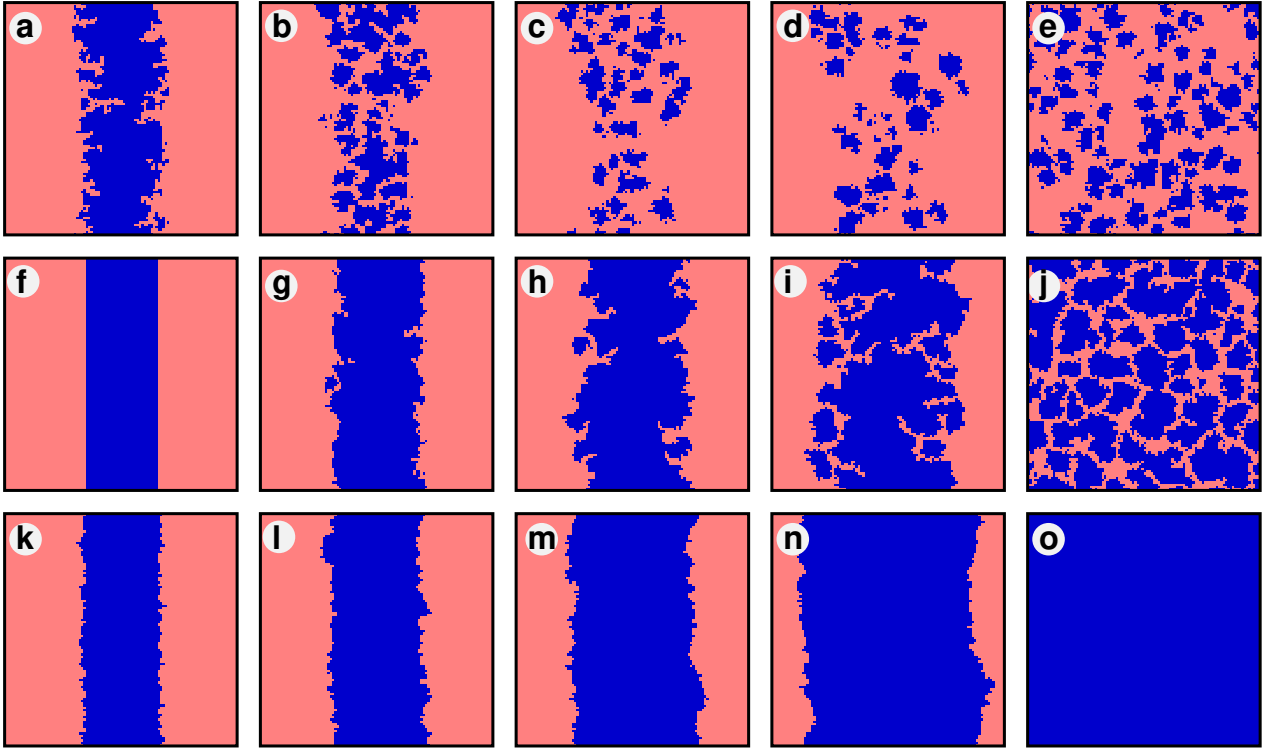,width=17.0cm}}
\caption{Evolution of the spatial distribution of strategies over time reveals that, even in the presence of equally effective antisocial rewarding, prosocial rewarding promotes the spatial selection for cooperation in the studied public goods game. Depicted are snapshots of the square lattice over time from left to right, as obtained for $r_2=1$ (top row), $r_2=1.3$ (middle row), and $r_2=2$ (bottom row). For clarity, we have used a prepared initial state for all cases with only a stripe of rewarding cooperators (blue) and rewarding defectors (pale red) initially present in the population, as depicted in panel (f). It can be observed that in the absence of rewarding (top row) the interface separating the two competing strategies is broken easily, and network reciprocity alone can ultimately sustain only small cooperative clusters. However, as the effectiveness of pool rewarding increases (middle and bottom row), the interface is strengthened, which makes the phalanx of cooperators more effective. The latter helps to reveal the
benefit of aggregated cooperators in structured populations. In all three cases the synergy factor for the main public goods game is $r_1=3.8$.}
\label{snapshots}
\end{figure*}

These results are different from those obtained with random initial conditions in well-mixed populations, and they are likely to appear contradictory because there is no obvious advantage given to cooperators over defectors as the value of $r_2$ increases. In fact, defectors benefit just as much given that they run an identical pool rewarding scheme as cooperators. So why is the evolution of cooperation promoted? The answer is rooted in the possible aggregation of cooperators, which can easily emerge spontaneously in a structured population. It is therefore instructive to monitor the evolution of the spatial distribution of strategies over time, as obtained for different values of $r_2$. Results are presented in Fig.~\ref{snapshots}, where for clarity we have used a prepared initial state with only a stripe of rewarding cooperators (blue) and rewarding defectors (pale red) initially present, as it is illustrated in panel (f). In all cases the synergy factor for the main public goods game was set to $r_1=3.8$.

The top row of Fig.~\ref{snapshots} shows the evolution obtained at $r_2=1$, which corresponds to the traditional, reward-free public goods game. It can be observed that the initially straight interface separating the two competing strategies disintegrates practically immediately. There is a very noticeable mixing of the two strategies, which ultimately helps defectors to occupy the larger part of the available space. Here cooperators are able to survive solely due to network reciprocity, but at such a relatively small value of $r_1$ only small cooperative clusters are sustainable. Nevertheless, we note that in a well-mixed population defectors would wipe out all cooperators at such a small value of the synergy factor.

Snapshots depicted in the middle row of Fig.~\ref{snapshots} were obtained at $r_2=1.3$, where thus both antisocial and prosocial pool rewarding mechanisms are at work. Here the final state is still a mixed $R_C+R_D$ phase (see also Fig.~\ref{phase}), but the fraction of cooperators is already significantly larger than in the absence of rewarding. Larger cooperative clusters are sustainable in the stationary state, which is due to an
augmented interfacial stability between competing domains. In addition to traditional network reciprocity, clearly the formation of more compact cooperative clusters is further promoted by the introduction of pool rewarding, and this despite the fact that both antisocial and prosocial rewarding mechanisms are equally strong.

If an even higher value of $r_2$ is applied, the interface that separates $R_C$ and $R_D$ players becomes impenetrable for defectors. The two strategies do not mix at all, which maintains the phalanx of cooperators \cite{nowak00}. Accordingly, the latter players simply spread into the region of defectors until they dominate completely. This scenario is demonstrated in the bottom row of Fig.~\ref{snapshots}, where the final stationary state is indeed a pure $R_C$ phase.

\begin{figure}
\centerline{\epsfig{file=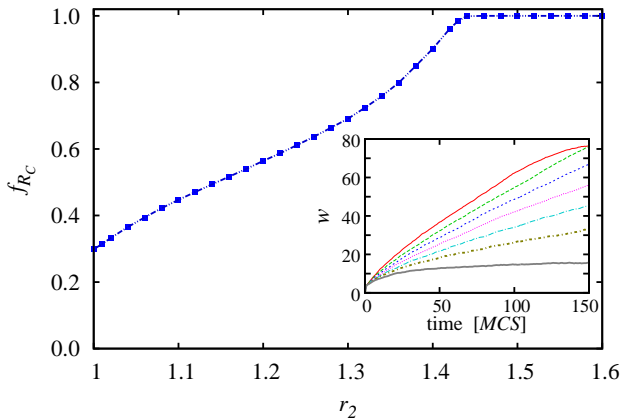,width=9cm}}
\caption{Quantitative evidence in support of enhanced spatial selection for cooperation in the studied spatial public goods game with antisocial and prosocial pool rewarding. The inset shows early stages of the evolution of the width $w$ of the mixed zone, where both strategies are present after initially starting from a prepared initial state, as depicted in panel~(f) of Fig.~\ref{snapshots}. From top to bottom the curves were obtained for $r_2=1$, $1.1$, $1.2$, $1.3$, $1.4$, $1.5$, and $2$, and they correspond to the average over 100 independent runs at system size $L \times L = 100 \times 100$. In all cases the synergy factor for the main public goods game is $r_1=3.8$. The main panel shows the corresponding increase in the fraction of rewarding cooperators $f_{R_C}$ as $r_2$ increases, thus indicating that the favourable outcome is indeed due to the enhanced stability of interfaces in structured populations. This enables cooperators to dominate completely even at low values of $r_1$, where in well-mixed populations they would not be able to survive, and where based only on network reciprocity they would fare poorly.}
\label{reward}
\end{figure}

As demonstrated in the middle and the bottom row of Fig.~\ref{snapshots}, the introduction of pool rewarding supports the aggregation of akin players and results in more stable interfaces between competing domains. This fact enhances the positive impact of network reciprocity further and provides an even more beneficial condition for cooperation. This favourable consequence of rewarding can be studied directly by monitoring how the width $w$ of the mixed zone -- the stripe where both strategies are present -- evolves over time when the evolution starts from the prepared initial state that is depicted in the panel (f) of Fig.~\ref{snapshots}. According to the definition of the width of the mixed zone, $w=0$ in panel~(j), while it becomes $w=L$ in panels~(e) and (j). The inset of Fig.~\ref{reward} shows how $w$ increases in time for different values of $r_2$ increasing from top to the bottom curve. Clearly, as the effectiveness of rewarding increases, the width of the mixed zone increases slower and slower. While for low values of $r_2$ the width of the mixed zone increases until eventually it covers the whole population (see panel (e) of Fig.~\ref{snapshots} for a demonstration), for sufficiently large values of $r_2$ the width remains finite, saturating and never exceeding a certain threshold. This result provides quantitative evidence that the interface between the two competing strategies remains intact, and that in fact the compact phalanx of cooperators cannot be broken by defectors. This in turn directly supports the evolution of cooperation to the point where defectors are wiped out completely, and this despite of the fact that they are able support each other by means of antisocial rewarding.

Based on the results presented thus far, it is possible to provide a clear rationale why a strategy-neutral intervention, like in this case the introduction of pool rewarding that at least in principle ought to benefit cooperators and defectors equally, is able to have such a biased impact on the final evolutionary outcome. In particular, pool rewarding yields an additional payoff to the players only if they aggregate and form at least partly uniform groups. This is beneficial for cooperators because it also helps them to obtain a competitive payoff from the original public goods game. In other words, the long-term benefits of cooperation come into full effect. The fate of defectors, on the other hand, is under this assumption entirely different. They can benefit from the antisocial rewarding scheme if they aggregate into uniform groups, but then they are unable to exploit the efforts of cooperators in the main public goods game. If they do not aggregate, then the benefits from antisocial rewarding become void. Either way, unlike cooperators, defectors are unable to enjoy the rewards as well as maintain a sustainable level of public goods. Ultimately, this favours the evolution of cooperation even though the intervention on the game is strategy neutral in that it does not favour one or the other strategy directly by granting it a higher payoff. This argument also explains why the same positive outcome is not attainable from a random initial state in well-mixed populations, where it was concluded that the possibility of antisocial rewarding utterly shatters any evolutionary benefits to cooperators that might be stemming from prosocial rewards \cite{dos-santos_m_prsb15}. If the interactions among players are well-mixed, then of course neither cooperators nor defectors can aggregate locally, which is a fundamental condition to reveal the long-term benefits of cooperation in a collective enterprise, even if the population contains strategies that seek to actively promote antisocial behaviour.

\begin{figure}
\centerline{\epsfig{file=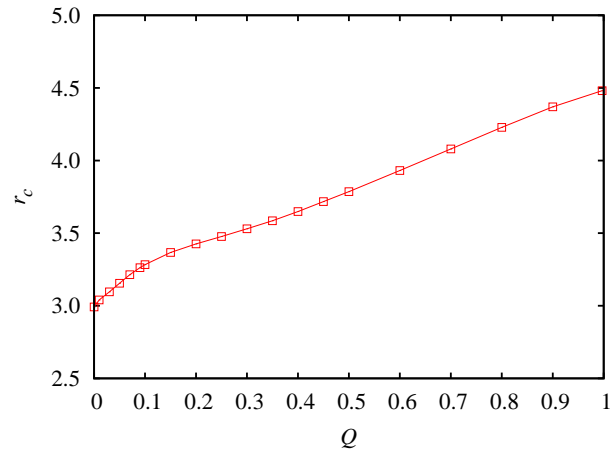,width=8.66cm}}
\caption{Random rewiring enables defectors to enjoy the benefits of antisocial rewarding and free-riding on the cooperative efforts of others, but relatively healthy conditions for the evolution of cooperation are maintained even if the randomness of the interaction network is high. Depicted is the critical value of $r_1=r_c$ at which the population arrives to a pure $R_D$ phase. It can be observed that $r_c$ values increase steadily as $Q$ increases, yet stay below the survival threshold of cooperators in a well-mixed population even at $Q=1$. The synergy factor for the antisocial and prosocial pool rewarding scheme is $r_2=2$. Qualitatively identical results are obtained also for other values of $r_2$.}
\label{mix}
\end{figure}

To corroborate our main arguments further, it is instructive to consider the studied spatial public goods game on alternative interaction networks, in particularly on such where random mixing can be controlled and adjusted deliberately. To that effect, we randomly rewire a certain fraction $Q$ of links that constitute the originally considered square lattice, so that for small values of $Q$ we obtain a regular small-world network, while in the $Q \to 1$ limit we obtain a regular random network, as described in \cite{szabo_jpa04}. Essentially, we thereby allow players to expand the range of their interactions to players that are well outside their local neighbourhood. In agreement with the above-outlined arguments, this randomness in the interaction structure ought to prevent defectors from suffering the negative consequences of aggregation with their like, thus allowing them to further exploit the cooperative efforts of others whilst still enjoying the benefits of antisocial pool rewarding. We note that at high values of $Q$ it is very likely that the direct neighbours of any given player are not strongly connected. The aggregation of players with the same strategies therefore looses effect. Defectors who are members in one group can also be members in completely different groups, where perhaps the exploitation of cooperators is still possible. We test this argument quantitatively in Fig.~\ref{mix}, where we show how the critical synergy factor $r_c$ of the main public goods game for which the population arrives to the pure $R_D$ phase increases as $Q$ increases. Indeed, as we increase the fraction of random links, more and more defectors are able to enjoy the benefits of antisocial rewarding as well as the benefits of free-riding on the cooperative efforts of others. As a countermeasure, a higher synergy factor is needed to prevent defectors from taking over. Nevertheless, even at $Q=1$ the required value of $r_1$ is still below the survival threshold of cooperators in a well-mixed population, and up to $Q=0.5$, when half of all the links are randomly rewired, there are still benefits to strategy-neutral pool rewarding that go beyond those offered solely by network reciprocity. We thus conclude that antisocial rewarding does not deter public cooperation in structured populations, even if the randomness of the interaction network is high. Detrimental effects of strategies that seek to promote antisocial behaviour appear to be significantly lessened if the assumption of a well-mixed population is replaced by a structured population.

\section*{4. Discussion}
We have studied the joint impact of antisocial and prosocial pool rewarding in a public goods game, in particular focusing on potential detrimental effects on the evolution of public cooperation that may stem from strategies that seek to actively promote antisocial behaviour. We have been motivated by the fact that strategies that promote antisocial behaviour are surprisingly common in human societies \cite{herrmann_s08} and in various inter-specific social systems \cite{wang_rw_e10}, as well as by the fact that recent research on a similar variant of the public goods game in a well-mixed population has shown that antisocial rewarding can lead to the breakdown of cooperation if the latter is contingent on pool rewarding \cite{dos-santos_m_prsb15}. By considering akin-like pool rewarding rather than peer rewarding, we also depart from the mainstream efforts to study the effects of rewards in structured populations \cite{szolnoki_epl10, szolnoki_njp12, szolnoki_prx13}, and join the recent \cite{henrich_s06, gurerk_s06, sigmund_n10, szolnoki_pre11, traulsen_prsb12, cressman_jtb12, perc_srep12, vasconcelos_ncc13, zhang_by_expecon14, sasaki_bl14} (and not so recent \cite{yamagishi_jpsp86}) trend in recognizing the importance of institutions for the delivery of positive and negative incentives to cooperate in collective enterprises.

Our research reveals that, in structured populations, the detrimental effects of antisocial rewarding are significantly more benign than in well-mixed populations. Even if the interaction network lacks local structure and has many long-range links, and in this sense approaches conditions that one might hope to adequately describe by a well-mixed population, antisocial rewarding still fails to upset the effectiveness of prosocial rewarding in promoting public cooperation. We have shown that the rationale behind this rather surprising result is rooted in spatial pattern formation, and in particular in the necessity of alike strategies to aggregate if they want to enjoy the benefits of rewarding. While this condition is actually beneficial for cooperators because it helps them to obtain a competitive payoff from the original public goods game, defectors suffer significantly because they are no longer able to free-ride on the cooperative efforts of others. The situation for defectors is thus a lot like Sophie's choice, in that they can either enjoy the benefits of antisocial rewarding or the benefits of free-riding on the public goods, but they can not do both simultaneously. And just one of the two options is not sufficient to grant them evolutionary superiority over cooperators. Therefore, even in the presence of antisocial rewarding, prosocial rewarding still offers benefits to cooperators that go well beyond network reciprocity alone.

An interesting alternative interpretation of the studied public goods game is to consider the introduction of antisocial and prosocial pool rewarding as a strategy-neutral interference on the original rules of the social dilemma \cite{perc_bs10}. We emphasize that neither defectors nor cooperators gain an obvious evolutionary advantage from the introduction of pool rewarding -- in fact, both strategies benefit exactly the same. It is therefore puzzling why, in the long run, cooperators turn out as the favoured strategy. This is in fact different from what was reported before for punishment, where available results indicate that antisocial punishment prevents the coevolution of punishment and cooperation \cite{rand_jtb10}, unless individuals have a reputation to lose \cite{hilbe_srep12}, or if individuals have the freedom to leave their group and become loners \cite{garcia_jtb12}. Nevertheless, the results presented in our study add to the favourable aspects that positive incentives to promote cooperation have over negative incentives \cite{dreber_n08, rand_s09}. The likely unwanted consequences of punishment are well know and include failure to lead to higher total earning, damage to reputation, and invitation to retaliation \cite{dreber_n08, herrmann_s08, rockenbach_pnas11}.

Summarizing, we have shown that antisocial rewarding does not necessarily deter public cooperation in structured populations, even if the randomness of the interaction network is high. This is because the delivery of rewards is contingent on the aggregation of alike strategies, which effectively prevents defectors from free-riding on the public goods. At the same time, the aggregation enhances the spatial selection for cooperation in evolutionary social dilemmas and thus helps to expose the long-term benefits of cooperative behaviour.

\begin{acknowledgments}
This research was supported by the Hungarian National Research Fund (Grant K-101490), the Slovenian Research Agency (Grant P5-0027), and by the Deanship of Scientific Research, King Abdulaziz University (Grant 76-130-35-HiCi).
\end{acknowledgments}

\end{document}